%% file: Manuscript.tex
\documentclass[aps,prl,twocolumn,showpacs,superscriptaddress,groupedaddress]{revtex4-1}
\usepackage[utf8x]{inputenc}
\usepackage{graphicx}
\usepackage{dcolumn}
\usepackage{bm}
\usepackage{amsmath}
\usepackage{amssymb}
\usepackage{float}
\usepackage{subfigure}
\usepackage{subfloat}
\usepackage{color}

\begin{document}
    
    \title{Implications of surface noise for the motional coherence of trapped ions}
    
    \author{I. Talukdar$^{1}$, D. J. Gorman$^{1}$, N. Daniilidis$^{1}$, P. Schindler$^{1}$, S. Ebadi$^{1,3}$, H. Kaufmann$^{1,4}$, T. Zhang$^{1,5}$, H. H\"{a}ffner$^{1,2}$}
    \email{hhaeffner@berkeley.edu}
    
    \affiliation{$^1$Department of Physics, University of California, Berkeley, 94720, Berkeley, CA}
    \affiliation{$^2$Materials Sciences Division, Lawrence Berkeley National Laboratory, Berkeley, CA $94720$}
    \affiliation{$^3$University of Toronto, Canada}
    \affiliation{$^4$University of Mainz, Germany}
    \affiliation{$^5$Peking University, China}

    \date{\today}

    \begin{abstract}
        Electric noise from metallic surfaces is a major obstacle towards quantum applications with trapped ions due to motional heating of the ions. Here, we discuss how the same noise source can also lead to pure dephasing of motional quantum states. The mechanism is particularly relevant at small ion-surface distances, thus imposing a new constraint on trap miniaturization. By means of a free induction decay experiment, we measure the dephasing time of the motion of a single ion trapped 50~$\mu$m above a Cu-Al surface.  From the dephasing times we extract the integrated noise below the secular frequency of the ion. We find that none of the most commonly discussed surface noise models for ion traps describes both, the observed heating as well as the measured dephasing, satisfactorily. Thus, our measurements provide a benchmark for future models for the electric noise emitted by metallic surfaces.

    \end{abstract}
    
    
    \maketitle
    
	Understanding decoherence constitutes an integral part in the development of any quantum technology. All present implementations of a quantum bit have to contend with the deleterious effects of decoherence \cite{Haeffner2008, Pashkin2009, Hayashi2003, Krojanski2004, Saffman2010}. It is usually identified with an irreversible loss of information from a quantum system when the system interacts with its environment.  Decoherence manifests itself in a decay of the phase relationships between energy-eigenstates, and can be characterized by the coherence time $T_{2}$, when these relationships decay to $1/e$ of their original values. This decay is usually thought to be a combination of two processes, a population relaxation with time constant $T_{1}$ and pure phase relaxation with a characteristic time of $T_{\phi}$.
	
	Trapped ion technology is one of the most promising candidate platforms to host a scalable quantum information processor. Before it can attain this goal, key challenges from decohering noise processes have to be overcome \cite{Brownnutt2014}. In particular, motional heating of ions trapped above room temperature microfabricated surface traps has been discussed as a serious roadblock as it limits the miniaturization of ion traps due to the increased heating as the ion is trapped closer to the trap electrodes.  It is now firmly established that the main obstacle in achieving low motional heating stems from electric field fluctuations emanating from the metallic surfaces \cite{Hite2012,Daniilidis2014,Bruzewicz2015}. Electric fields resonant with the secular frequency of the trapped ion excite the ion's secular motion and thus lead to motional heating. Some models attribute the electric field noise to fluctuating dipole-like sources on the trap electrode surfaces \cite{Daniilidis2011, Safavi-Naini2011,Safavi-Naini2013} while others to fluctuating potential patches \cite{Dubessy2009,Low2011} or surface diffusion of adatoms \cite{Gesley1985,Wineland1998,Hite2012}. Electric field noise from surfaces has been found to span across a wide range of distance and frequency regimes, impacting diverse fields, including scanning probe microscopy \cite{Hofer2003}, gravitational wave experiments \cite{Pollack2008}, superconducting electronics \cite{Pashkin2009}, detection of Casimir forces \cite{Kim2010a}, and studies of non-contact friction \cite{Stipe2001}. Investigation of the spectral characteristics of this noise has therefore received considerable attention.

	\begin{figure}[H]
		\begin{center}
			\includegraphics[width = 0.4\textwidth]{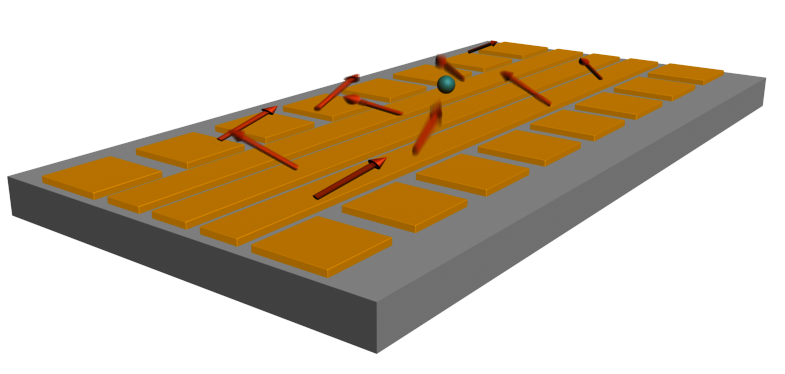}      
			\caption{(Color online) Schematic of fluctuating dipoles on a microfabricated surface trap causing motional heating and dephasing.}
			
			\label{fig:schematic}
		\end{center}
	\end{figure}  
		    
	Studies on trapped ions have so far primarily focused on motional heating due to the noise source’s electric field. For the fidelity of entangling operations, however, the T$_2$ time, rather than the heating time, is the more appropriate figure of merit. In particular, gradients of a noisy electric field will result in random changes of the instantaneous secular frequencies of the ion motion, leading to loss of phase coherence of the ion's motion without heating. More importantly, in such a case, it is not the noise resonant with the secular frequency that is relevant. Rather, the total noise below the secular frequency can cause potentially significant decoherence as the noise spectral density typically increases towards lower frequencies.  Here we discuss this mechanism and how to measure it.
	      
	To illustrate this point we consider noise caused by fluctuating dipoles.  We note that the following arguments could made to any model where the individual noise sources are small as compared to the ion-surface distance. In particular, the arguments apply to models of patches with fluctuating dipole moments \cite{Dubessy2009,Low2011} or surface diffusion of adatoms \cite{Gesley1985,Wineland1998,Hite2012}. Expanding the potential $\phi = \frac{\vec{\mu}.\vec{r}_0}{|r_0|^3}$ at the ion position $\vec{r} = (0,d,0)$ due to a surface dipole $\vec{\mu}(t)$ to the second order, we get
	\begin{equation}\label{eq:potential}
	\begin{aligned}
	\phi \simeq \frac{\mu}{4\pi\epsilon_0}[\frac{3dx_d}{r_0^5}x+\frac{x_d^2-2d^2+z_d^2}{r_0^5}(y-d) + \\
	\frac{3d(-4x_d^2+d^2+z_d^2)}{2r_0^7}x^2  +\frac{3d(2d^2-3(x_d^2+z_d^2))}{r_0^7}(y-d)^2\\
	+\frac{3x_d(x_d^2-4d^2+z_d^2)}{r_0^7}x(y-d)] + \mathcal{O}(r^3)
	\end{aligned}
	\end{equation}
	
	Here, $r_0=\sqrt{x_d^2+d^2+z_d^2}$ is the distance between the ion and the dipole at $(x_d,0,z_d)$, and we have only kept terms linear and quadratic in $y$ and one horizontal direction $x$. For a homogeneously distributed layer of dipoles on a plane with surface density $\sigma_d$, the electric field fluctuations along a direction parallel to the trap surface will have an autocorrelation
	\begin{align}\label{eq:Ex}
	\langle E_x(t)E_x(0)\rangle
	&= \int_{\textrm{surface}} \sigma_{d}\langle \mu(t)\mu(0)\rangle \left(\frac{3dx_d}{r_0^5}\right)^2 \textrm{d}A \nonumber \\
	&= \frac{3\pi\sigma_d}{8(4\pi \epsilon_0)^2d^{4}}\langle \mu(t)\mu(0)\rangle
	\end{align}
	This will give rise to electric field noise with a spectral density \cite{Turchette2000} $S_{E_x}(\omega)=2\int_{-\infty}^{\infty}\langle E_x(t)E_x(0)\rangle e^{i\omega t}\textrm{d}t$.
	
	Additionally, fluctuations of the quadrupole field $Q(t) = \frac{3d(-4x_d^2+d^2+z_d^2)}{4\pi\epsilon_0 2r_0^7}\mu(t)$ along $x$ will cause the secular frequency of the ion motion to fluctuate in time. The autocorrelation will be $\langle Q(t)Q(0)\rangle = \frac{45\pi}{32(4\pi\epsilon_0)^2 d^6}\langle \mu(t)\mu(0)\rangle$. Thus, the noise spectral density of the quadratic potential fluctuations, $S_{Q_{xx}}=\frac{15}{4d^2}S_{E_x}$, increases rapidly with $1/d^6$ as the ion is brought close to the surface.
	
	It is desirable to place the ions near the surface for performing faster ion shuttling operations as well as implementing state manipulation strategies integrated in the chip itself \cite{Ospelkaus2011, AudeCraik2013}. However, at reduced ion-surface distances, besides higher heating, pure dephasing might start to play an increasingly important part in the total decoherence. Furthermore, dephasing is particularly sensitive to low frequency noise. In combination with the fact that virtually nothing is known about the size of the electric field noise in ion traps at low frequencies ($\leqslant$ 100 kHz), there is a need to characterize the pure dephasing $T_\phi$ in addition to the motional heating $T_1$. Here, we find this time using a free induction decay experiment.

\begin{figure}
		\begin{center}
			\includegraphics[width = 0.4\textwidth]{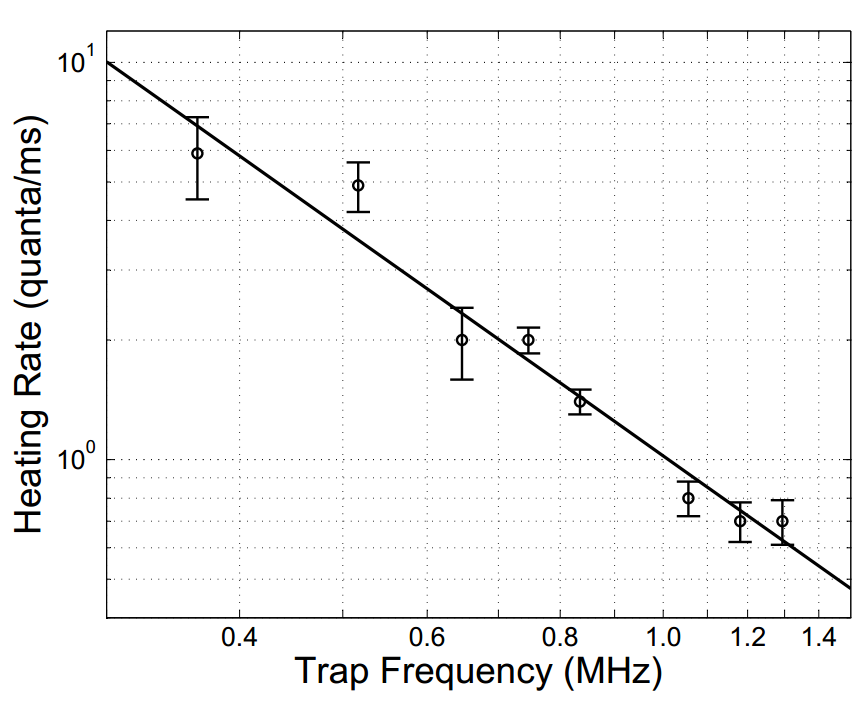}
			\caption{Measured motional heating rates at different axial trap frequencies (open circles) shows a scaling of $\omega^{-1.9(0.2)}$ (solid line).} \label{fig:heatingrate}
		\end{center}
	\end{figure}

	For our experiments, we use an Al-Cu surface trap microfabricated with a process identical to that used in Ref.~\cite{Daniilidis2014}. This surface trap was not subjected to any sputter treatment. Radio-frequency voltages create trapping potentials at a location 50~$\mu$m above the surface with secular frequencies 1.5~MHz, and 1.2~MHz along the radial directions while static voltages provide adjustable confinement along the axial direction between 0.37~MHz and 1.3~MHz. After laser cooling the ion, we measure its final temperature along the axial direction by observing Rabi oscillations on the $S_{1/2}$$\rightarrow$$D_{5/2}$ transition.  Specifically, we estimate the temperature, expressed in terms of the average population, $\bar{n}$, of the axial mode of the ion's motion, from the decay of Rabi oscillations \cite{Wineland1998}. By inserting various delays between the laser cooling and temperature readout steps we can infer the motional heating rate $\dot{\bar{n}}$ of the ion. Fig.~\ref{fig:heatingrate} shows measured heating rates ranging from 6(1)~quanta/ms to 0.7(1)~quanta/ms for trap frequencies between 0.365~MHz and 1.3~MHz. Thus, a typical $T_{1}(=1/\dot{\bar{n}})$ time is 1.0 ms at 1.0~MHz. The measurements show a scaling with the trapping frequency as $\dot{\bar{n}}\propto \omega^{-1.9(0.2)}$ or $S_E\propto \omega^{-0.9(0.1)}$ since the electric field noise density depends on the heating rate as \cite{Turchette2000} $S_E=\frac{4m\hbar \omega}{q^2}\dot{\bar{n}}$, where $q$ is the charge of an ion of mass $m$.

	To study solely the dephasing of the ion's motion, we create displaced thermal states of the harmonic oscillator \cite{Meekhof1996}. For this, we apply a train of pulses of light resonant with the cooling $S_{1/2}$$\rightarrow$$P_{1/2}$ transition to the laser-cooled ion at a pulse rate close to the trap frequency, detuned by $\delta_m$. By modulating the radiation pressure on the ion, this predominantly performs a displacement operation $\hat{D}(\alpha)$ on the Doppler-cooled state thereby creating a displaced thermal state $\rho_{\rm{disp}} = \hat{D}(\alpha)\rho_{\rm{dopp}}\hat{D^{\dagger}}(\alpha)$ \cite{Saito1996}. After a certain delay $\tau$, the relative phase between the ion's motion and the pulse train will have evolved by $\delta_m\tau$. A second displacement pulse train can be applied to probe the coherence of the phase. If there is no external perturbation of the ion's motion, the amplitude of the ion motion would oscillate sinusoidally as a function of time between $2\alpha$ ($\delta_m\tau=0,2\pi$) and $0$ ($\delta_m\tau=\pi$) with unity contrast.  In the presence of any dephasing noise, random changes to the phase of the ion's motion will lead to oscillations with a reduced contrast.
	
	After the two displacement operations, we determine the ion's motional amplitude by driving the red sideband $|S_{1/2},n \rangle$$\rightarrow$$|D_{5/2},n-1\rangle$. For an excitation pulse of length $t$, the probability of the ion to be excited to the $|D\rangle$ level is
	\begin{equation} \label{probD}
	P_D(t)=\frac{1}{2}\Big[1-\sum_n\rho_{n,n}\cos(\Omega_{n,n-1}t) \Big]
	\end{equation}
	The excitation thus depends on the coupling strength to the sideband $\Omega_{n,n-1}$$=$$ \sqrt{n}\eta\Omega$, and the occupation probability $\rho_{n,n}=(\frac{1}{\bar{n}+1})(\frac{\bar{n}}{\bar{n}+1})^n e^{-\frac{|\alpha|^2}{\bar{n}+1}}L_n(-\frac{|\alpha|^2}{\bar{n}(\bar{n}+1)})$ of the $n^{\rm{th}}$ harmonic oscillator state \cite{Saito1996}. Here, $\eta$ is the Lamb-Dicke parameter, $\Omega$ the coupling strength of the carrier transition, and $L_n$ the Laguerre polynomial of the $n^{\textrm{th}}$ degree\cite{Saito1996}. Therefore, by measuring $P_D$, we can deduce the displacement $|\alpha|$ \cite{Ramm2014}.
	\begin{figure}
	    \begin{center}
	        \includegraphics[width = 0.35\textwidth]{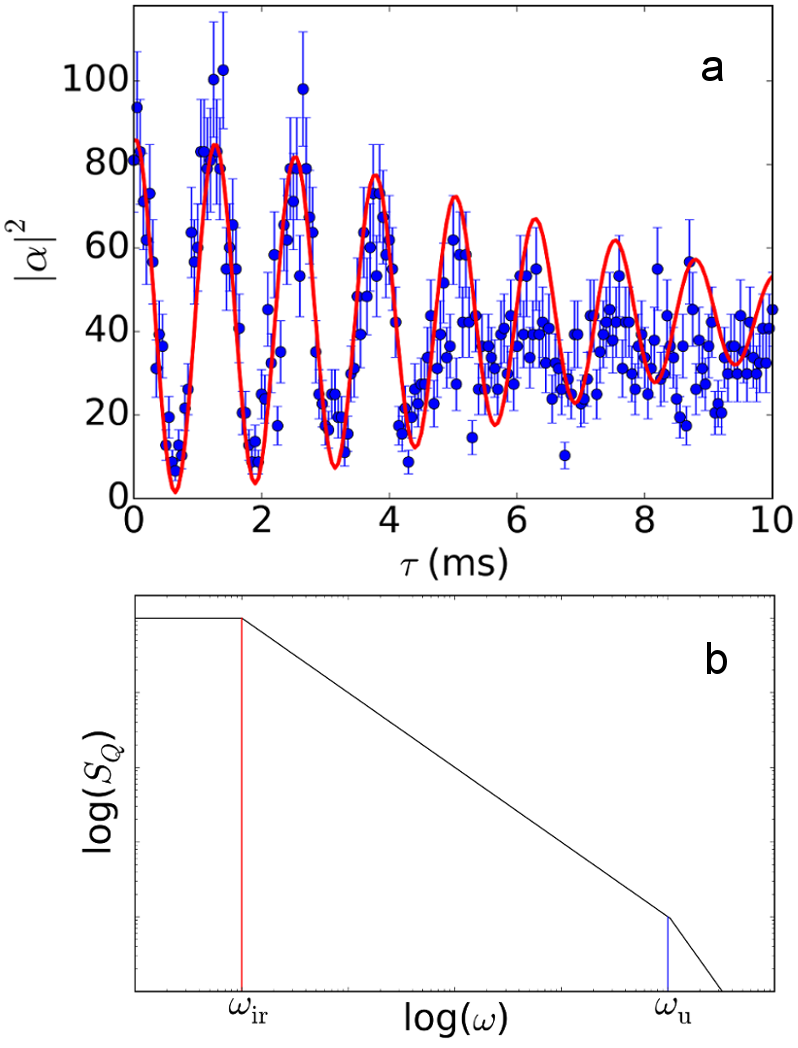}        
	        \caption{ (Color online) (a) Decay of the amplitude of a displaced thermal state of motion. Solid line is a fit of the measured data (filled circles) to dephasing under the influence of noise with a spectrum as depicted in (b). The error bars denote statistical errors from 200 measurements for each point. The spectrum rolls off from a frequency independent part to a $1/\omega^{\beta}$ dependent region at a variable low cut-off frequency $\omega_{\textrm{ir}}$ till a high cut-off frequency $\omega_{\textrm{u}}$.}
	        \label{fig:dephase}        
	    \end{center}
	\end{figure}
	The result of an experiment performed at a trapping frequency of 880~kHz is shown in Fig.~\ref{fig:dephase}. We produce states with large displacements, typically $\bar{n}=|\alpha|^2$$\sim$50, such that we can neglect population changes $\Delta\bar{n}/\bar{n}$ due to motional heating on the order of a few quanta/ms.
	Using this method we observe a decay of the amplitude of oscillations with a $1/e$ time constant of $T_{\phi}=$~3.9(0.5)~ms.
	    
	 Theoretical models suggest that the surface noise scales with frequency as  $S_E\propto 1/\omega^{\beta}$. Experimental measurements have found  $\beta$ to range from\cite{Brownnutt2014,Allcock2011} 0.4  to 1.6. $1/\omega^{\beta}$-noise is ubiquitous in many physical systems \cite{Dutta1981}. It is well understood that in order to avoid a divergence of the noise power, $\int_{0}^{\infty}S(\omega)\textrm{d}\omega\rightarrow \infty$, a low-frequency cut-off $\omega_\textrm{ir}$ for such noise must exist. Here we estimate the cut-off frequency from our measurements of the dephasing of the motional states. In the context of trapped-ion quantum computing, knowledge of the cut-off allows one to put a bound on the relevant time scale of dephasing processes and gain insight into microscopic models of the noise.
    
    As shown in Refs. \cite{Cywinski2008, Uhrig2011}  
    the coherence of the motional state can be shown to evolve as $e^{-\chi(\tau)}$. The decay term $\chi(\tau) = \int_{0}^{\infty}S(\omega)F(\omega)\textrm{d}\omega$ describes a noise spectral density $S(\omega)$ probed by a filter function $F(\omega)$. To relate $S(\omega)$ to the quadrupole noise we note that the oscillator-environment interaction is governed by the Hamiltonian
    \begin{equation}\label{eq:hamil}
    \hat{H}=\hbar[\omega_0 + \eta(t)]\hat{a}^{\dagger}\hat{a}.
    \end{equation}
    $\omega_0$ is the trap frequency, $\hat{a}^{\dagger}(\hat{a})$ is the oscillator mode creation (annihilation) operator and, $\eta(t)=qQ(t)x^2$ is the random fluctuation of the quadrupole potential energy.
    The fluctuations lead to
    \begin{align}\label{eq:noise_spec}
    S(\omega)
    &=\frac{1}{2\pi}\int_{-\infty}^{\infty}\langle \eta(t)\eta(0)\rangle e^{-i\omega t}\textrm{d}t \nonumber \\
    &= \frac{1}{2\pi}\bigg(\frac{q}{m\omega_0}\bigg)^2\int_{-\infty}^{\infty}\langle Q(t)Q(0)\rangle e^{-i\omega t}\textrm{d}t \nonumber \\
    &= \bigg(\frac{q}{2\sqrt{\pi}m\omega_0}\bigg)^2 S_{Q}.
    \end{align}
    The spectral filtering in our experiment is due to a free induction decay of duration $\tau$. Therefore, the filter function $F(\omega)=\frac{1}{2}\frac{\sin^2(\omega\tau/2)}{(\omega/2)^2}$ is given by the Fourier transform of the free induction decay period.  Thus, the motional state amplitude evolves as,
    \begin{equation}\label{eq:alphaT}
    |\alpha|^2(\tau)
    = \cos(\delta_m\tau)\exp[-\left(  \frac{q}{2\sqrt{\pi}m\omega_0}\right)^2\int_{0}^{\infty}S_Q(\omega)F(\omega)\textrm{d}\omega ]
    \end{equation}
    
    We consider a noise spectrum that rolls over from a flat region at low frequencies to a $1/\omega^{\beta}$ region at a cut-off frequency $\omega_\textrm{ir}$.  This is shown schematically in Fig.~\ref{fig:dephase}(b). Such a spectrum is motivated by the idea of activated random processes with relaxation rates following a log-uniform distribution between $\omega_\textrm{ir}$ and a high frequency cut-off $\omega_\textrm{u}$ \cite{Dutta1981}. With the filter function $F(\omega)$ dropping to zero quickly past the delay time $\tau$, the experiment is sensitive only to low frequency noise.  Therefore, we ignore the high frequency cut-off in our estimate of the noise spectrum. Thus, the integral in the exponent of Eq.~(\ref{eq:alphaT}) comprises of two parts,
    \begin{equation}\label{eq:spectrum}
    \int\limits_{0}^{\infty}\rightarrow A[\int\limits_{0}^{\omega_\textrm{ir}}F(\omega)\textrm{d}\omega + \int\limits_{\omega_\textrm{ir}}^{\infty}\bigg(\frac{\omega_\textrm{ir}}{\omega}\bigg)^{\beta}F(\omega)\textrm{d}\omega ].
    \end{equation}
    \begin{figure}[H]
        \begin{center}
            \includegraphics[width = 0.4\textwidth]{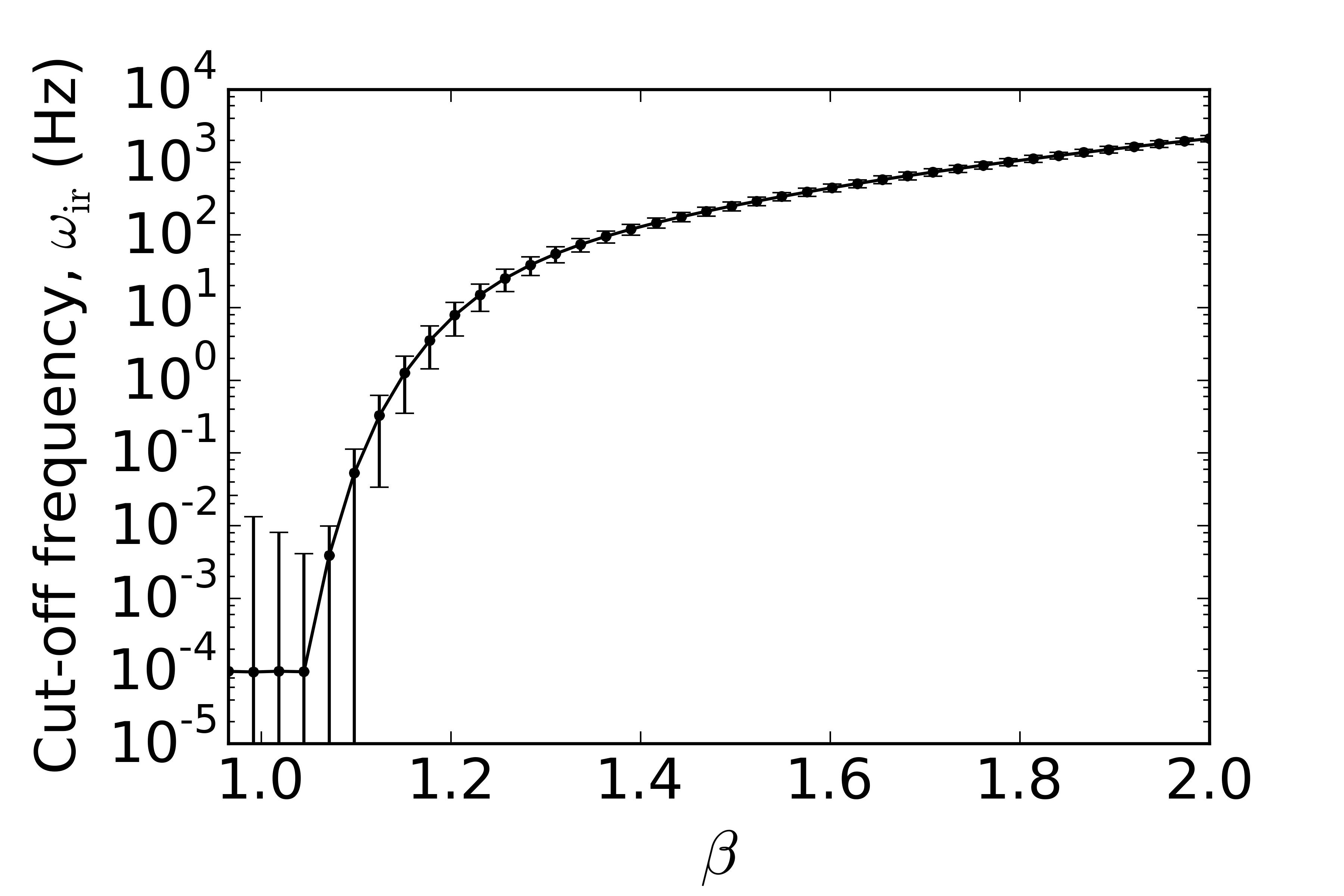}        
            \caption{Estimates of the lower frequency cutoff, $\omega_{ir}$ for different exponents of the noise frequency scaling. The error bars represent standard errors of the best fit values.}
            \label{fig:cutoff}        
        \end{center}
    \end{figure}
    Here, $A\propto\dot{\bar{n}}$ is a constant dependent on the heating rate. We performed a numerical fit of dephasing under the influence of this spectrum to the experimental data of Fig.~\ref{fig:dephase} for a range of $\beta$, with $\omega_\textrm{ir}$ as a free parameter, and scaled with respect to the observed heating rate of $\dot{\bar{n}}$=1 quanta/ms at 1~MHz. As seen in Fig.~\ref{fig:cutoff}, the low frequency cut-off depends on the specific heating rate frequency exponent $\beta$. The fact that the estimated low frequency cut-offs do not reduce further below $\beta \approx 1$ can be attributed to the minimum sensitivity of the experiment. The entire experiment takes around a few tens of minutes to complete, and therefore the slowest frequency fluctuations to which it is sensitive to are of order $10^{-4}$ Hz. To enforce this in the analysis we bound the lowest obtainable low cut-off by this value.  
    
    The observed dephasing can also originate from a relative instability of the trapping potentials. We have measured the rms fluctuations in our electrode voltage supply sources to be $\Delta V = 10 \mu$V within a measurement frequency range of 2 mHz to 100 kHz. To estimate the the effect of this on the trap frequencies we note that the potential due to a surface trap can be written as $U(x,y,z) = \sum_{j}\sum_{i}M_{ji}V_{i}Y_{j}$. Here, $M_{ji}$ is the coefficient of a multipole expansion of the potential at the ion position  due to a voltage $V_{i}$ on the $i^{\mathrm{th}}$ electrode, and $Y_{j}$ is the multipole component \cite{Schindler2015}. Therefore, voltage fluctuations can change the trap frequencies as $\Delta\omega/\omega = \Delta U/2U = \Delta V\sum_{i}M_{ji}/(2\sum_{i}M_{ji}V_{i})$. For the quadrupole coefficients and voltages used in the experiment, the dephasing time would have been $T_{\phi}=1/\Delta\omega=$ 36 ms. Thus, electrode voltages should not be limiting the phase stability of the ion's motion.
    
    In conclusion, we have investigated the effect of electric quadrupole noise emanating from surfaces of planar ion traps on the motion of an ion. To probe for pure motional dephasing of an ion trapped 50~$\mu$m above a surface, we performed a free induction decay experiment with a displaced thermal state. The measured phase coherence $T_\phi = 4$~ms is longer than the population relaxation time $T_1$ of 1~ms. Assuming $T_1\propto d^4$ and $T_{\phi}\propto d^6$, we expect for similar surfaces $T_{1}=T_\phi$ when the ion-surface distance reaches $50~\mu\textrm{m}\times\sqrt{{\mathcal{T}_1}/{\mathcal{T}_{\phi}}} = 25~\mu$m  (here, ${\mathcal{T}_1}$ and ${\mathcal{T}_{\phi}}$ are the measured values of the respective quantities). This $d^6$ scaling of the dephasing time will reduce the achievable entangling gate fidelities in smaller ion trap structures, imposing a practical limit to the extent to which ion traps can be miniaturized for quantum computing applications.
    
    Various models have been proposed to explain excessive surface
    noise \cite{Dubessy2009,Low2011,Daniilidis2011,Kumph2015,
    Safavi-Naini2011,Safavi-Naini2013, Gesley1985, Wineland1998}. As of yet it is unclear which of those are relevant, or if any of them are universally applicable. Learning more about the low frequency noise can constrain or validate some of those models. For example, the surface diffusion model \cite{Gesley1985,Wineland1998} predicts an exponent of $\beta\simeq 1.5$ with an estimated low frequency cut-off on the order of $10^{-7}$~Hz. Assuming $\beta = 1.5$ in the frequency regime below 100~kHz, we extract from our data a cut-off of $\sim300$~Hz, much higher than the cut-off estimated by the same model. Thus, this surface diffusion model might need a refinement to describe the noise of the surface studied consistently. Another model suggests adatoms of high molecular mass bound to the surface as the source for noise \cite{Safavi-Naini2011}. The estimated cut-off of this model is between 1-10~MHz, and scales inversely with the atomic mass of the adatom. This model would therefore require very large weakly bound atomic masses adsorbed to the surface to explain the observed dephasing in Fig.~\ref{fig:dephase}. Thus, both models describe our data unsatisfactory and it is very likely that some other noise process either of technical or physical nature is responsible for the dephasing, i.e. low frequency decoherence. We have carefully analysed possible dephasing due to voltage noise and find it not sufficient either. However, it is very difficult to exclude with certainty technical noise sources and thus to ascertain that the low frequency decoherence actually stems from surface noise.
    
    In the future, one can hope for more direct evidence for surface induced dephasing by combining the dephasing measurement technique described here along with surface modification methods. For instance, both the noise as well as the cut-off of the model presented in Ref.~\cite{Safavi-Naini2011} and its extensions \cite{Safavi-Naini2013} increase at higher temperatures. Thus changes in the dephasing rate as a function of the
    temperature will not only prove the existence of surface induced dephasing but will also give important information as to which mechanisms lead to surface noise and limit the coherence of the ion motion. On a more general level, electric field noise near surfaces is an ubiquitous theme throughout science and engineering. Hence establishing
    accurate models of surface noise will have a major impact across many disciplines in science and technology.
    
    P.S. was supported by the Austrian Science Foundation (FWF) Erwin Schr\"odinger
    Stipendium 3600-N27. This research was partially funded by the
    Office of the Director of National Intelligence (ODNI), Intelligence
    Advanced Research Projects Activity (IARPA), and through the Army Research
    Office grant W911NF-10-1-0284. All statements of fact, opinion or
    conclusions contained herein are those of the authors and should not
    be construed as representing the official views or policies of IARPA,
    the ODNI, or the U.S. Government.
    
    \input{References.bbl}

\end{document}

%% file: References.bbl
%